\documentclass{emulateapj}
\usepackage{apjfonts}
\usepackage{epsf}
\begin{document}

\slugcomment{}
\shortauthors{J. M. Miller et al.}
\shorttitle{Revisiting Cool Disks in ULXs}

\title{Revisiting Putative Cool Accretion Disks in Ultraluminous X-ray Sources}

\author{J.~M.~Miller\altaffilmark{1},
        D.~J.~Walton\altaffilmark{2},
        A.~L.~King\altaffilmark{1},
        M.~T.~Reynolds\altaffilmark{1},
        A.~C.~Fabian\altaffilmark{3},
        M.~C.~Miller\altaffilmark{4},
        R.~C.~Reis\altaffilmark{1}}

\altaffiltext{1}{Department of Astronomy, University of Michigan, 500
Church Street, Ann Arbor, MI 48109-1042, jonmm@umich.edu}

\altaffiltext{2}{Cahill Center for Astronomy \& Astrophysics,
  California Institute of Technology, Pasadena, CA 91125}

\altaffiltext{3}{Institute of Astronomy, University of Cambridge,
  Madingley Road, Cambridge, CB3 OHA, UK}

\altaffiltext{4}{Department of Astronomy, University of Maryland,
  College Park, MD, 20742, USA}

\keywords{accretion disks, black hole physics}

\label{firstpage}

\begin{abstract}
Soft, potentially thermal spectral components observed in some ULXs
can be fit with models for emission from cool, optically-thick
accretion disks.  If that description is correct, the low temperatures
that are observed imply accretion onto ``intermediate-mass'' black
holes.  Subsequent work has found that these components may follow an
inverse relationship between luminosity and temperature, implying a
non--blackbody origin for this emission.  We have re-analyzed numerous
{\it XMM-Newton} spectra of extreme ULXs.  Crucially, observations
wherein the source fell on a chip gap were excluded owing to their
uncertain flux calibration, and the neutral column density along the
line of sight to a given source was jointly determined by multiple
spectra.  The luminosity of the soft component is found to be
positively correlated with temperature, and to be broadly consistent
with $L \propto T^{4}$ in the measured band pass, as per blackbody
emission from a standard thin disk.  These results are nominally
consistent with accretion onto black holes with masses above the range
currently known in Galactic X-ray binaries, though there are important
caveats.  Emission from inhomogeneous or super-Eddington disks may
also be consistent with the data.
\end{abstract}

\section{Introduction}
Ultra-luminous X-ray sources (ULXs) are variable, off-nuclear X-ray
sources in nearby galaxies, with luminosities in excess of $L \simeq
10^{39}~ {\rm erg}~{\rm s}^{-1}$ (the Eddington luminosity of an $M =
10~ M_{\odot}$ black hole).  Though they are luminous, most ULXs are
likely familiar objects.  The well-known source GRS 1915$+$105, for
instance, can by modestly super-Eddington for its mass, depending on
the model and energy band considered.  The vast majority of ULXs only
exceed $L \simeq 10^{39}~ {\rm erg}~{\rm s}^{-1}$ by a small margin
(Swartz et al.\ 2011; for a recent review, see Feng \& Soria 2011).

The small subset of ULXs with luminosities of $L \simeq 10^{40}~ {\rm
  erg}~ {\rm s}^{-1}$ and above are potentially more interesting, as
they might be powered by accretion onto so-called
``intermediate-mass'' black holes (IMBHs), or an accretion flow that
has genuinely defeated the isotropic Eddington limit.  Of course, this
small sub-class could be a combination of these two phenomena.  An
alternative possibility is that the emission from such sources is
actually anisotropic, perhaps owing to a ``funnel'' in the inner
accretion disk (King et al.\ 2001).

{\it XMM-Newton} has revolutionized studies of ULXs, making it
possible to obtain spectra that require multiple components.  Early
efforts to decompose the best ULX spectra found evidence of separate
soft and hard components.  The soft components could be fit with disk
models, and low temperatures obtained -- generally $kT =$0.2--0.3~keV
-- provided some evidence of accretion onto IMBHs since $T \propto
M_{BH}^{-1/4}$ and $kT = 1$~keV is typical for accretion onto
stellar-mass black holes close to the Eddington limit (e.g. Miller et
al.\ 2003; Miller, Fabian, \& Miller 2004).  This spectral
decomposition is based on a close analogy with better-known X-ray
binaries and may not be unique nor appropriate for ULXs.  However,
subsequent studies of soft component variability found that these
would-be disks may not be blackbody-like (Kajava \& Poutanen 2009;
Feng \& Kaaret 2007, 2009; also see Soria 2007).  Other recent work
has discovered a spectral roll-over above the Fe K band (Gladstone,
Roberts, \& Done 2009).  This can be interpreted as evidence of
super-Eddington accretion, though the spectra still require
independent soft components with low characteristic temperatures.  In
this work, we examine the nature of the best ULX spectra from nearby
sources with $L \simeq 10^{40}~ {\rm erg}~ {\rm s}^{-1}$, factoring in
recent studies that impact the possibility of winds and variable
absorption.

\section{Sample and Data Reduction}
We considered the sample of nearby $L \simeq 10^{40}$~erg/s sources
observed with {\it XMM-Newton} and reported in Miller, Fabian, \&
Miller (2004), Gladstone, Roberts, \& Done (2009), and Kajava \&
Poutanen (2009).  ULXs at distances greater than 5~Mpc were not
considered owing to their reduced photon flux.  Of the remaining
sources, NGC 1313 X-2 was not considered as any evidence that it
exceeds $L = 10^{40}$~erg/s appears weak and highly model--dependent,
and evidence for a cool disk component is only marginal (e.g. Miller
et al.\ 2003).  IC 342 X-1 -- which is near to $L = 10^{40}$~erg/s --
was excluded owing to the modest significance of a putative cool
component.  In exploratory fits, the column density along the line of
sight to IC 342 X-1 fell in the $N_{H} = 0.5-1.5\times 10^{22}~ {\rm
  cm}^{-2}$ range, greatly complicating the detection of $kT \simeq
0.2$~keV emission.  With these constraints, our sample includes
Holmberg IX X-1, NGC 1313 X-1, Ho II X-1, NGC 5408 X-1, and NGC 5204
X-1.  All archival {\it XMM-Newton} observations of these sources were
downloaded and reduced using SAS version 12.0.1.

The EPIC-pn camera has the highest collecting area across the full
0.3--10.0~keV band, and is best calibrated for spectral fitting.  For
simplicity and self-consistency, our analysis was restricted to
spectra obtained using the EPIC-pn camera.  Unlike {\it Chandra}, {\it
  XMM-Newton} does not dither, and when a source lands within the gap
between chips in EPIC cameras, its effective exposure and encircled
energy fraction can be affected, ultimately compromising flux
estimates.  Observations wherein the source image fell within chip
gaps were therefore rejected, in order to ensure robust flux
determinations.

Background regions were extracted on the same CCD as the source, and
generally using an extraction region of the same size.  The background
regions were analyzed to identify periods of background flaring, and
to create a GTI file to exclude these intervals when extracting events
for spectral analysis.  Spectra, backgrounds, and responses were then
calculated using the appropriate tools.  All spectra were grouped to
require at least 25 counts per bin using the FTOOL ``grppha'', in
order to ensure the validity of $\chi^{2}$ statistical tests.

\section{Analysis and Results}
All spectra were fit using XSPEC version 12.2 (Arnaud et al.\ 1996).
Neutral interstellar absorption was fit using the ``tbabs'' model.  As
required by ``tbabs'', the ``vern'' atomic cross-sections and ``wilm''
elemental abundances were used.  The Milky Way's contributions to the
total neutral column density along these lines of sight are small, and
we therefore used a single ``tbabs'' component to account for both
Galactic absorption and the column within each ULX host galaxy.  Solar
abundances were assumed in all fits (but see the Discussion).

The spectral resolution afforded by dispersive spectrometers has
recently been leveraged to address the extent to which absorption may
drive spectral evolution in accreting systems (Miller, Cackett, \&
Reis 2009).  The depth of individual photoelectric absorption edges
remains remarkably constant across a broad range in luminosity in
low-mass X-ray binaries, in binaries with potential
``intermediate-mass'' stars such as Cygnus X-2, and even in Cygnus X-1
(which accretes from an O 9.7 Iab supergiant).  This argues that the
line of sight column density should be held constant in spectral fits.
Current limits on Fe K emission and absorption lines in ULXs are
commensurate with detections in Galactic X-ray binaries, and far below
expections if line strengths scale with the mass accretion rate
(Walton et al.\ 2012, 2013).  We therefore fit all EPIC-pn spectra of
a given ULX jointly, such that the interstellar column density was
jointly determined and had a common value for every spectrum.

For simplicity, we chose to fit the soft, potentially thermal
components with the well-known ``diskbb'' model (Mitsuda et
al.\ 1984).  To characterize the hard flux in each ULX spectrum, we
used the ``compTT'' model (Titarchuk 1994).  The use of ``compTT'' is
important for characterizing the turn--over seen in the 6--10~keV band
in many sensitive spectra of ULXs with $L \simeq 10^{40}$~erg/s
(e.g. Gladstone et al.\ 2009; Walton et al.\ 2013).  It produced
statistically superior fits to simple power-law models for the hard
flux.  The temperature of the low-energy thermal distribution $T_0$
was linked to that of the disk component.  The other crucial parameters
within ``compTT'', the electron temperature $kT_e$ and the optical
depth $\tau$ were allowed to vary.  However, few spectra are able to
constrain both parameters, and in those cases a value of $kT_e = 2.0$~keV
was adopted.  This value is broadly consistent with the spectra where
constraints were possible, and also broadly consistent with the values
reported in fits with ``compTT'' reported by Gladstone et al.\ (2009)
and Feng \& Kaaret (2009).

An additional diffuse emission component is present in the spectra of
NGC 5408 X-1, likely due to local warm gas and star formation.
Emission localized around 1 keV can be modeled with a ``mekal'' plasma
with $kT = 0.87(2)$~keV and a normalization of $8.0\pm0.5 \times
10^{-5}$.  These values were determined through joint fits.  This
correctly accounts for diffuse emission with a constant flux.

The results of our spectral fits are given in Table 1.  The procedure
of jointly determining the column density is clearly one that allows
for excellent fits.  In all cases, the joint fit returns a reduced
$\chi^{2}$ statistic that is close to unity.  The use of the
``compTT'' component with a turn-over above the Fe K band also
contributed to the excellent fits.  Figure 1 plots the X-ray
luminosity measured in the putative disk components, versus their
color temperature values.  Both in Table 1 and Figure 1, the
luminosity is restricted to the band in which the flux was actually
observed (0.3--10.0~keV).  

Figure 1 also plots the relationship expected for simple blackbody
emission, with a number of different normalizations.  It is
immediately apparent that the ULX putative disk components show a
clear, positive relationship between luminosity and temperature that
is qualitatively consistent with $L \propto T^{4}$.  (Note that band
pass effects can be important in effective luminosity versus
temperature relationships; it is possible that $L \propto T^{5}$ may be
anticipated for very cool disks measured in the 0.3--10.0~keV band.)
The lone exception is NGC 5408 X-1; the points from that source are
tightly clustered.

In order to quantify the apparent relationships, we ran statistical
correlation tests on the data from each individual source, and also
made simple least-squares fits to determine the slope of the data.
The results of these tests and fits are listed in Table 2.  Errors on
both luminosity and temperature were considered in fitting the data
and estimating errors on the slope.  Strong positive correlations are
confirmed in each source, apart from NGC 5408 X-1.  The data from Ho
IX X-1 and Ho II X-1 are formally consistent with $L \propto T^{4}$,
and the slope of NGC 1313 X-1 is consistent within 1.5$\sigma$.  The
slope of NGC 5204 X-1 is somewhat flatter.  The best traces of $L$ versus
$kT$ in stellar-mass black hole disks -- made using {\it Swift},
spanning three orders of magnitude in $L$, and also obtained using
``diskbb'' and ``compTT'' continuum components -- find more shallow
slopes, e.g. $L \propto T^{3.3\pm 0.1}$ (Rykoff et al.\ 2007, also see
Reynolds \& Miller 2013, Salvesen et al.\ 2013).

The errors on the slopes listed in Table 2 are large owing to the small
number of points available.  Similarly, the strongest correlations are
only significant at the 99\% level of confidence.  However, taken
literally, the data would suggest that the putative cool thermal disk
components in this small subset of extreme ULXs may indeed represent
disk emission.  Since disk tempertures this low are only seen in
stellar-mass black holes at or below $0.01~L_{Edd.}$, the results
would nominally indicate accretion onto more massive black holes,
since $T \propto M_{BH}^{-1/4}$ for black hole accretion.

\section{Discussion and Conclusions}
We have examined the relationship between the color temperature of
putative cool, thermal components versus their luminosity, in the
0.3--10.0~keV spectra of ULXs with $L_{tot} \simeq 10^{40}$~erg/s.  We
find evidence of positive correlations between the luminosity of these
putative disk components, and their color temperature.  In the
observed band, some temperature and luminosity trends are formally
consistent with the $L \propto T^{4}$ relationship expected for simple
blackbody emission from a standard thin accretion disk.  In all but
one source, the data are consistent with a slightly flatter
relationship that is observed in stellar-mass black holes in the same
band (e.g. $L \propto T^{3.3\pm 0.1}$, Rykoff et al.\ 2007; also see
Reynolds \& Miller 2013, Salvesen et al.\ 2013).  The lone exception
is NGC 5408 X-1, for which the data span a very small range in
luminosity and temperature, and thus no trend can be discerned.  Taken
literally, these results may support an interpretation of these
components as emission from cool accretion disks around
intermediate-mass black holes.  We therefore proceed to make a
critical examination of our methods and assumptions in this section.

As noted previously, some recent work has found putative disk
luminosity and temperature to be anti--correlated when fitting ULX
spectra with disk components at low energy (e.g. Feng \& Kaaret 2007,
2009; Kajava \& Poutanen 2009).  Our analysis employed a more
restrictive selection criterion, in that observations wherein the
source image landed on a chip gap were excluded.  Observations with
spurious or uncertain flux measurements are therefore omitted.
Moreover, the correlation we have found is specific: the {\it disk}
luminosity and the {\it disk} temperature are positively correlated.
We do not find strong correlations between {\it total} luminosity and
disk temperature.  Though coronae and disks must be linked, the need
for magnetic processes (see, e.g., Merloni \& Fabian 2002) means that
the luminosity of these components can be decoupled at times.

Our analysis also differs from prior efforts in that the column
density along the line of sight to a given source was jointly
determined by the numerous spectra, and not allowed to vary between
them.  It is possible that this method is faulty, especially
if ULXs are fueled by massive stars with variable, clumpy winds.
However, existing data appear to justify our assumptions and fitting
methods:

First, grating spectra of even Cygnus X-1 find a consistent column
across the binary phase, except perhaps when the O star is closest to
our line of sight (Miller, Cackett, \& Reis 2009).  Second, existing
spectra of NGC 1313 X-1 and Holberg IX X-1 now place extremely
restrictive limits on emission and absorption features in these
sources (Walton et al.\ 2013).  Any emission lines must have lower
equivalent widths than the lines seen in Galactic X-ray binaries with
massive companions.  This indicates that companion winds are largely
absent or very highly ionized.  Either way, companion winds are
unlikely to contribute to an evolving neutral column density.  Limits
on absorption lines are also below those detected from disk winds in
stellar-mass black holes and many AGN, again limiting the scope for
variable absorption (Walton et al.\ 2012, 2013; Pasham \& Strohmayer
2013).  Last, where dips are detected in ULXs (e.g. NGC 5408 X-1,
Grise et al.\ 2013), they are apparently quasi--periodic (Pasham \&
Strohmayer 2013), and thus inconsistent with clumps in a companion
wind.  Dips in Cygnus X-1 are clustered at $\phi = 0.95$ and $\phi =
0.6$, with those at $\phi = 0.6$ likely due to the accretion stream
impacting the outer disk; dips owing to clumps in the companion wind
appear to be spread randomly in phase (Balucinksa-Church et
al.\ 2000).  And, as noted previously, even in Cygnus X-1, the optical
depth in various edges is remarkably constant (e.g. Miller, Cackett,
\& Reis 2009).

In our treatment of the neutral column, we also assumed solar
abundances for all elements.  Some studies of the dwarfs in this
sample (in particular) have reported sub--solar metallicity values in
the ISM of those galaxies (e.g. Guseva et al.\ 2011, Egorov,
Lozinskaya, \& Moiseev 2013), and caution is warranted.  Fits to the
best X-ray spectra, however -- including gratings spectra obtained
with the RGS -- find abundances consistent with solar (Winter,
Mushotzky, \& Reynolds 2007; Pintore \& Zampieri 2012).  The same
effect is found in our data: if the absorption model for Ho II X-1 in
Table 1 is altered so that the metal abundances are only 10\% of
solar, a significantly worse fit is achieved ($\Delta \chi^{2} = 44$).
The same effect holds for the large spiral NGC 1313 ($\Delta \chi^{2}
= 132$).  Such results are not driven by contributions from the
Galactic column; in all cases, the Galactic column (as estimated by
Dickey \& Lockman 1990) is only a fraction of that measured.
Moreover, it is important to remember a basic facet of the observed
absorption edge features: one can trade intrinsic column density and
abundance, but their product has to match the observed edge depth.
Shifting abundance values will affect all spectra from a given source
in the same manner.

The spectral model we employed also has difficulties.  The ``comptt''
model includes thermal emission, and it therefore competes with the
external ``diskbb'' component for the soft X-ray flux.  In this sense,
it is not perfectly self-consistent.  However, ``comptt'' is required
to describe the roll-over in the 5--10 keV band (e.g. Gladstone,
Roberts, \& Done 2009), and this is now a standard model.  Alternative
Comptonization treatments, such as ``simpl'', only produce a power-law
and thus miss the observed roll-over.  For the fits presented in Table
1, the additional soft component is required by the data at extremely
high statistical significance.

There are important caveats, but our results nominally support the
possibility that soft components in ULX spectra represent emission from
standard accretion disks extending to the ISCO.  It is possible to
derive some simple mass estimates by scaling from stellar-mass black
holes: $M_{ULX} \simeq (M_{XRB} / M_{\odot}) \times
(T_{XRB}/T_{ULX})^{4}$, where $T_{XRB}$ is the disk temperature
typical for stellar-mass black holes in X-ray binaries close to the
Eddington limit (or, in a state analogous to that in which ULXs
accrete).  If we take $kT_{XRB} = 1$~keV, $M_{XRB} = 10~ M_{\odot}$,
$T_{ULX} = 0.2$~keV, then a mass of $M_{ULX} = 6250~ M_{\odot}$ is
implied.  However, not all transients that cycle through each of the
canonical spectral states are observed to have maximum temperatures of
$kT \simeq 1$~keV.  XTE J1650$-$500, for instance, had a maximum
temperature of $kT \simeq 0.6$~keV (e.g. Reis et al.\ 2013).  For
$kT_{XRB} = 0.5$~keV, $M_{XRB} = 5~ M_{\odot}$, and $kT_{ULX} =
0.25$~keV, a more modest mass of $M_{ULX} \simeq 80 M_{\odot}$ is
implied.  Clearly, the mass estimate is strongly dependent upon the
``typical'' temperature of a standard X-ray binary disk at Eddington.

These mass estimates are simplified.  There is no certainty that ULXs
radiate at Eddington.  Moreover, the data allow for flatter
relationships between luminosity and temperature, potentially
suggestive of a changing disk radius.  In this circumstance, and in
situations where the coronal energy drains the disk, mass estimates
are more complex, and generally imply lower masses ($M_{ULX} \leq 100~
M_{\odot}$, Soria 2007).  The putative cool disk components in our
models only emit a fraction of their luminosity in the observed
(0.3--10.0~keV) band.  However, these disks would dominate the source
luminosity in the 0.01--10.0~keV band, and would then be more
analogous with the disk--dominated ``high/soft''states seen in
stellar-mass black holes.  Depending on numerous details, the slopes
measured when considering bolometric disk luminosity versus
temperature may be 0.5--1.0 flatter than the values reported in Table
2.  This effect is within the error ranges quoted in the existing
fits, and assumes that the spectral model can safely be extrapolated
to lower energy values.

It is possible that the observed temperatures and flux trends could
represent emission from a locally--inhomogeneous disk, although the
observed temperature contrast is slightly greater than that envisioned
in current treatments (e.g. Dexter \& Quataert 2012).  It is also
possible that the putative cool disk components we have studied do not
originate close to the black hole, but rather outside of some
transition radius, with an hot, inner, super-Eddington disk
represented by the Comptonized component.  In this scenario, however,
considerable fine--tuning would likely be required for the different
sources and transition radii to create the observed temperatures and
trends.  Future monitoring campaigns aimed at improved traces of soft
component variability, and very deep observations that search for the
outflows expected in super-Eddington regimes, may help to better
understand this subset of ULXs.

\begin{figure}
\includegraphics[scale=0.42]{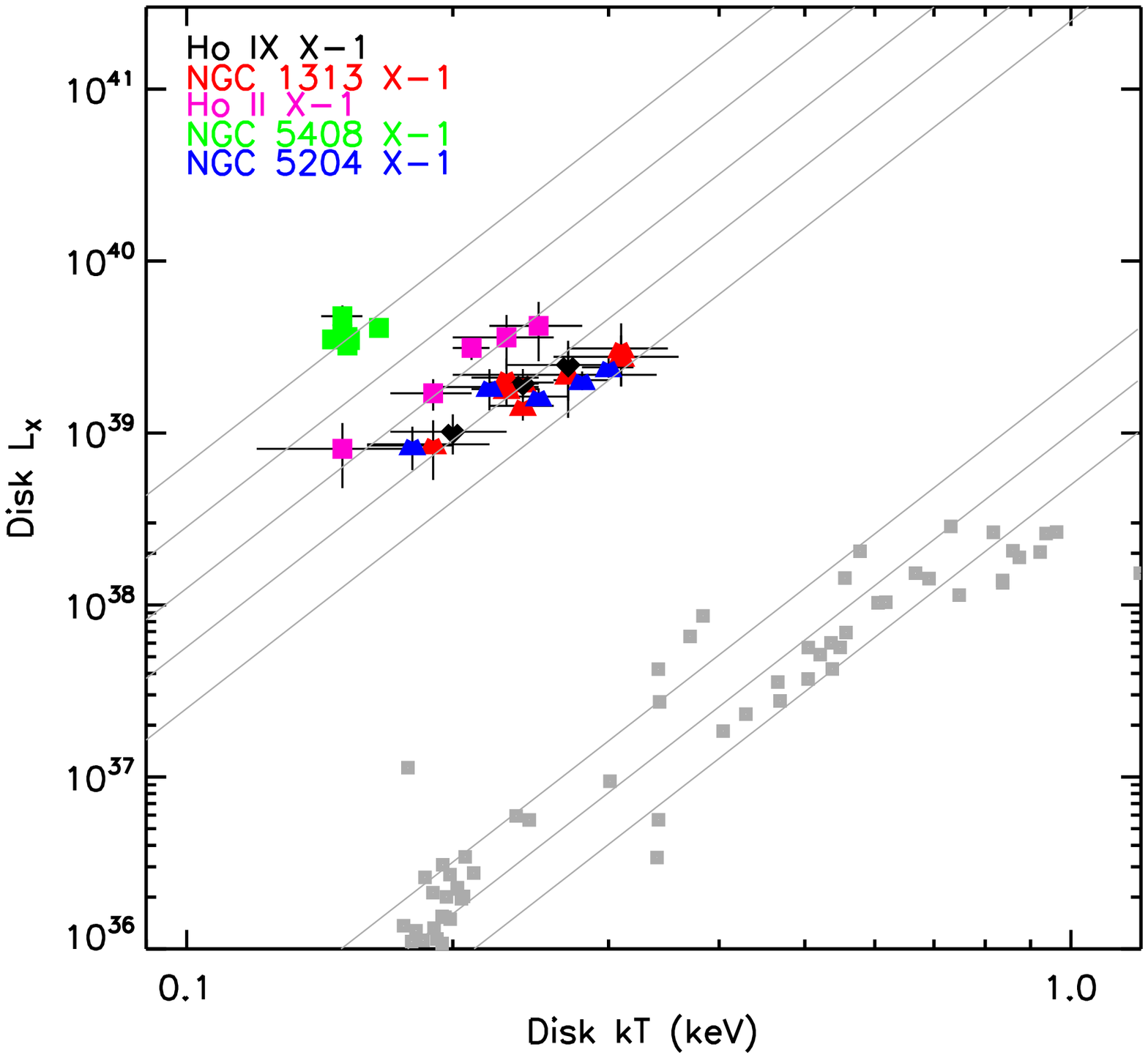}
\includegraphics[scale=0.42]{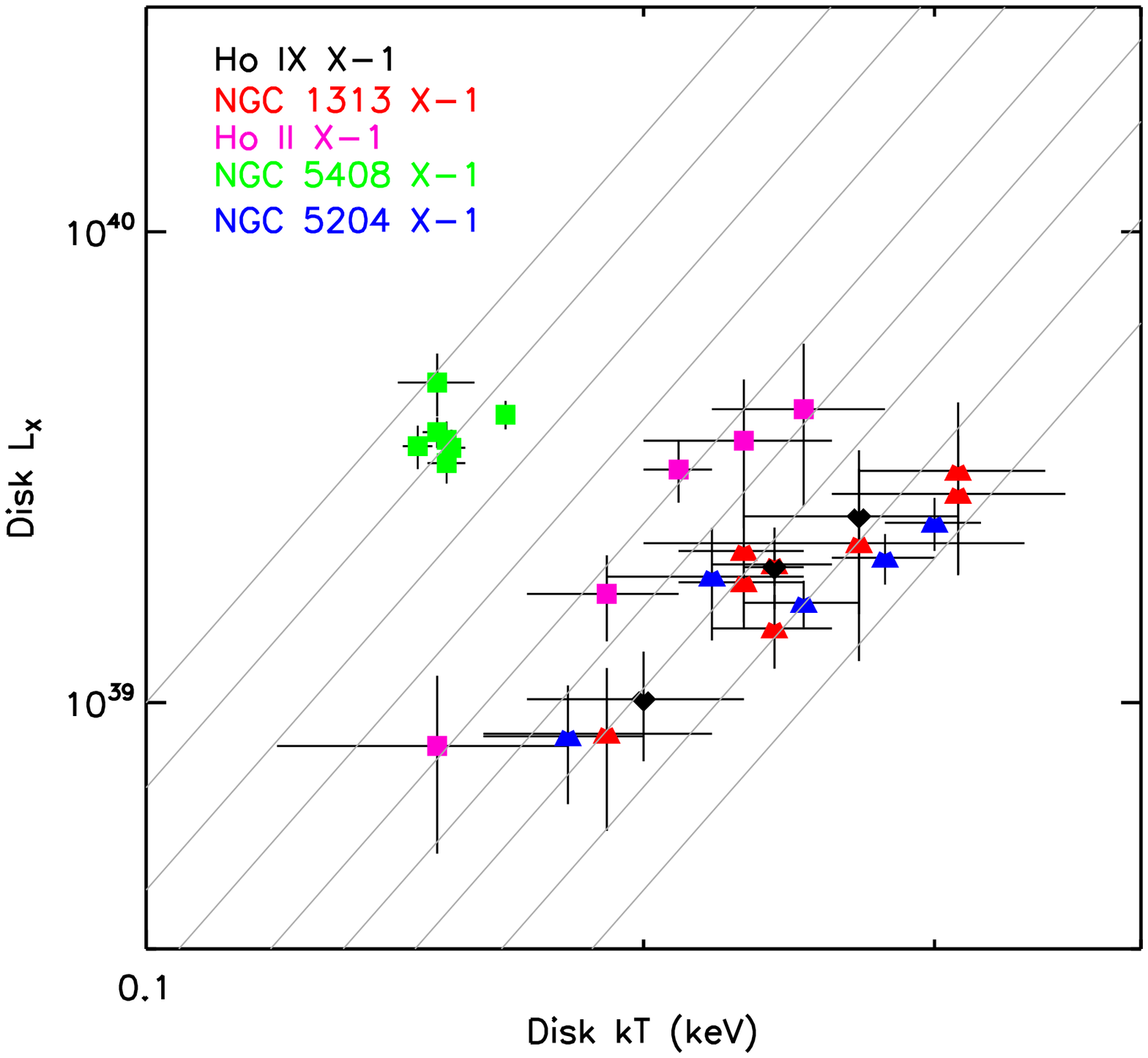}
\figcaption[t]{\footnotesize The figure above shows the luminosity of
  putative cool disk components versus their apparent temperature, as
  measured in numerous spectra of ULXs with a total luminosity
  frequently in excess of $L \simeq 10^{40}$~erg/s.  For comparison,
  data from a survey of stellar-mass black holes with {\it Swift}
  (Reynolds \& Miller 2013; LMC X-3, GRS 1915$+$105, GX 339$-$4, XTE
  J1752$-$223, and XTE J1817$-$330 are shown) are plotted in gray
  in the left-hand panel.  The diagonal gray lines depict $L \propto
  T^{4}$ with different normalizations.  As these sources are not
  expected to have exactly the same mass, there is no reason to expect
  that they should follow $L \propto T^{4}$ with a common
  normalization.  The left-hand and right-hand panels show the same
  ULX data; the right-hand panel merely examines a narrower range in
  $L$ and $kT$.}
\end{figure}
\medskip

\begin{table}[t]
\caption{Spectral Fitting Results}
\begin{tiny}
\begin{center}
\begin{tabular}{lllllllllll}
\tableline
Source & ObsID & $N_{H}$ & $kT$ & Norm. & $kT_e$ & $\tau$ & Norm & $L_{tot}$ & $L_{disk}$ & $\chi^{2}/\nu$ \\
 ~ &     ~  & ($10^{22}~ {\rm cm}^{-2}$) & (keV) & ~  & (keV) & ~ & ($10^{-4}$)  &  ($10^{40}~ {\rm erg}~ {\rm s}^{-1}$) & ($10^{39}~ {\rm erg}~ {\rm s}^{-1}$) & ~ \\
\tableline
Ho IX X-1 &  0112521001  &  0.157(7) &	0.27(4)	    &  20(7)   &	2.8(5)  &    7.2(9)  &	7.6(9)  &   1.3(5)	&  2.5(9)	 &	2139.9/2140 \\
~ &   0112521101         &	~    &      0.20(3) &	28(6)  &	5.7(2)	&    4.4(2)  &	6.1(4) 	&   1.4(4)	&  1.0(3)	 &	~  \\
 ~ &  0200980101	 &      ~    &      0.24(1) &	26(2) & 	2.45(8)	&    8.9(3)  &	7.2(2)	&   1.1(1)	&  1.9(2)        &	~ \\
\tableline
NGC 1313 X-1 &	0106860101 &	0.272(8) &  0.24(2)  &	28(5)	&  2.3(3)  &	8.5(7)	&  	4.2(3)  &   0.6(1)	  &  2.0(4)	  	 &    2513.8/ 2398 \\
  ~  &	        0150280201 &  ~	 &	    0.24(2)  &	19(3)	&  2.0	   &	10(3)	&	6(3)	&   0.5(1)	  &  1.4(3)	  	 & ~ \\	
  ~  &	        0150280601 &  ~	 &	    0.31(4)  &	16(6)	&  2.0	   &	7.5(5)	&	7(1)	&   0.9(3)	  &  3(1)	     & ~ \\
  ~  & 	        0150281101 &  ~	 &	    0.27(7)  &	17(6)	&  2.0	   &	6.6(3)	&	10(4)	&   0.9(3)	  &  2(1)	     & ~ \\
  ~  &	        0205230201 &  ~	 &	    0.31(5)  &	14(4)   &  2.0     &	9(1)   	&	4(1)	&   0.7(2)	  &  2.8(9)	     & ~ \\
  ~  &	        0205230401 &  ~	 &	    0.19(3)  &	32(9)  &  2.0	   &	4.2(2)	&	10(3)	&   0.5(1)	  &  0.8(3)	     & ~ \\
  ~  &	        0205230601 &  ~	 &	    0.23(3)  &	32(9)   &  2.0	   &	8.7(4)	&	5.4(5)	&   0.7(2)	  &  2.1(5)	     & ~ \\
  ~  & 	        0405090101 &  ~	 &	    0.23(2)  &	27(5)	&  2.2(2)  &	8.5(5)	&	4.3(1)	&   0.6(1)	  &  1.8(3)	  	 & ~ \\	
\tableline
Ho II X-1 & 0112520601	&  0.080(3) &	0.25(3)	&	70(25)	&	2.0 &	5.9(3)	&	12(3)	&   1.1(4)   &   4(2)   &      2033.6/1990 \\
 ~   &   0112520701     &   ~       &   0.23(3)	&	60(20)	&	2.0 &	6.6(3)	&	11(3)	&   1.0(4)   &   4(1)   &  ~ \\
 ~  &	 0112520901     &   ~       &   0.15(3)	&	140(50)	&	2.0 &	4.8(3)	&	6(3)	&   0.3(1)   &   0.8(3)   & ~ \\
 ~  &	 0200470101     &   ~       &   0.21(1)	&	85(12)	&	2.7 &	4.6(6)	&	13(1)	&   1.2(2)   &   3.1(5)   & ~ \\
 ~  &	 0561580401     &   ~       &   0.19(2)	&	84(15)	&	2.0 &	5.6(3)	&	6.4(6)	&   0.46(9)  &   1.7(4)   & ~ \\
\tableline
NGC 5408 X-1 &	0112290501  &  0.105(3)3 &	0.150(8)	&	350(50)	&	2.0	&	5.0(3)	&	6(1)	&	1.0(2)  &    4.8(7)	        &    3686.4/3551 \\
 ~ &		0302900101  &    ~       &	0.165(3)	&	177(12)	&	2.0	&	5.1(2)	&	4.1(3) 	&       0.72(7)  &   4.1(3)	      &   ~ \\
 ~ &		0500750101  &  ~         &	0.152(4)	&	220(20) &	2.0	&	5.57(9)	&	4.9(3)	&	0.67(7)  &   3.2(3)	    & ~ \\
~ &	        0653380201  &  ~         &	0.150(3)	&	280(20) &	2.0	&	5.5(1)	&	6.5(3)	&	0.83(7)  &   3.8(3)	    & ~ \\
 ~ &		0653380301  &  ~	 &      0.146(3)	&	290(30)	&	2.0	&	5.30(5)	&	6.9(4)	&	0.8(1)  &    3.5(4)	     & ~ \\
 ~ &		0653380401  &  ~	 &      0.152(3)	&	242(22)	&	2.0	&	5.42(6)	&	5.7(3)	&	0.78(8)  &   3.6(4)	     & ~ \\
 ~ &		0653380501  &  ~	&       0.153(3)	&	220(20)	&	2.0	&	5.58(6)	&	5.5(2)	&	0.75(7)  &   3.5(3)	     & ~ \\
\tableline
NGC 5204 X-1 &  0142770101  &  0.049(4)	&	0.18(2)	 &	27(7)	&	2.0 &	7.6(3)	&	3.2(5)	&       0.4(1)   &   0.9(2) & 1634.4/1601 \\
 ~           &	0142770301  &	~	&        0.28(2) &	9.0(9)  &	2.0 &	8.1(8)	&	2.3(4)	&	0.5(1)   &   2.0(3) & ~ \\
 ~           &	0405690101  &   ~	&	0.21(3)	 &	27(6)   &	2.0 &	5.8(3)	&	6(1)	&	0.7(2)   &   1.8(5) & ~ \\
 ~           &	0405690201  &   ~	&	0.30(2)	 &	9(1)    &	2.0 &	6.8(3)	&	2.8(4)	&	0.6(1)   &   2.4(3) & ~ \\
 ~           &	0405690501  &   ~	&	0.25(2)	 &	12(1)   &	2.0 &	7.7(3)	&	2.7(3)	&	0.47(7)  &   1.6(2) & ~ \\
\tableline
\end{tabular}
\vspace*{\baselineskip}~\\ \end{center} 
\tablecomments{The table above lists the results of joint spectral
  fits to the sources and observations in our sample, over the
  0.3--10.0 keV range.  The column density $N_{H}$ was jointly
  determined.  Where errors are not given, the parameter was fixed
  (see the text).  In calculating luminosity values we assumed
  distances of 3.6~Mpc, 3.7 Mpc, 3.4 Mpc, 4.8~Mpc, and 4.3~Mpc for Ho
  IX X-1, NGC 1313 X-1, Ho II X-1, NGC 5408 X-1, and NGC 5204 X-1,
  respectively (Paturel et al.\ 2002, Tully 1988, Karachentsev et
  al.\ 2002, Karachentsev et al.\ 2002, Tully 1988).}
\vspace{-1.0\baselineskip}
\end{tiny}
\end{table}
\medskip

\begin{table}[t]
\caption{Correlation Tests and Fits}
\begin{footnotesize}
\begin{center}
\begin{tabular}{llllll}
\tableline
Source & $\rho$ & $P_{FA}$ & $\tau$ & $P_{FA}$ & index \\
\tableline
Ho IX X-1 & 1.0 & 0.0 & 1.0 & 0.117 & $3.2\pm 2.6$  \\
Ho II X-1 &  1.0 & 0.0 & 1.0 & 0.014 & $3.7\pm 1.8$ \\
NGC 1313 X-1 & 0.84 & 0.0096 & 0.718 & 0.0129 & $2.4\pm 1.0$ \\
NGC 5408 X-1 &  -0.09 & 0.846 & -0.05 & 0.8745 & $1.7\pm 0.9$ \\
NGC 5204 X-1 & 0.9 & 0.037 & 0.8 & 0.05 & $1.9\pm 0.6$ \\
\tableline
\end{tabular}
\vspace*{\baselineskip}~\\ \end{center} 
\tablecomments{The table above lists the results of correlation tests
  of putative disk temperature $kT$ and luminosity $L_{disk}$, for the
  values given in Table 1.  The Spearkman's rank correlation
  coefficient $\rho$ and Kendall's $\tau$ coefficients and their
  associated false alarm probabilities (probability of false
  correlation) are given.  The index reported in the last
  column is the slope obtained in least-squares fits to the $L_{disk}$
  and $kT$ data from each individual source.  For simple blackbody
  emission, $L \propto T^{4}$ is expected, corresponding to an index
  of 4.0 in the final column.}
\vspace{-1.0\baselineskip}
\end{footnotesize}
\end{table}
\medskip


\end{document}